\documentclass[aps,prl,twocolumn,showpacs,preprintnumbers,amsmath,amssymb,superscriptaddress,floatfix]{revtex4}
\usepackage{graphicx}
\usepackage{dcolumn}
\usepackage{bm}
\usepackage{stmaryrd}
\usepackage{latexsym}
\usepackage{amssymb}
\usepackage{amsfonts}
\usepackage{amsmath}
\usepackage{fancybox}
\usepackage{color}

\definecolor{MyDarkBlue}{rgb}{0,0.08,0.45}
\begin{document}

\title{Nano-indentation of circular graphene flakes}

\author{Mehdi Neek-Amal }
\email{neek@nano.ipm.ac.ir}
\affiliation{School of Physics, Institute for Research in Fundamental Sciences, IPM 19395-5531 Tehran, Iran}
\affiliation { Department of physics, Shahid Rajaei University, Tehran 16788, Iran}
\author{Reza Asgari }
\email{asgari@theory.ipm.ac.ir}
\affiliation{School of Physics, Institute for Research in Fundamental Sciences, IPM 19395-5531 Tehran, Iran}
\begin{abstract}
Nano-indentation of circular graphene flakes are studied by using Molecular Dynamics simulation.
We show that the theory of continuum elasticity based on
nonlinear F\"{o}pple-Hencky equations
is applicable on a circular suspended graphene flake. Our simulation results are plenty compatible to those results calculated by the nonlinear elasticity theory. We find the force-displacement curves in good
agreement with the recent experimental measurements and conclude they are temperature independent.
In addition, we find the vibration frequency for such a system and monitor that it behaves as a sinusoidal manner at small circular graphene size.
\end{abstract}

\maketitle

\noindent
{\em Introduction}---
Understanding of the mechanical properties of graphene, a newly realized two-dimensional electron
system~\cite{novoselov,geim} is crucial in guiding its growth and applications. Booth \emph{et al} \cite{Giem2008} have
measured mechanical properties in a series of experiments on
a suspended graphene and showed that graphene has an extraordinary
stiffness which can supports an additional weight of many
crystalline copper nano-particles.
A nonlinear behavior of
stress-strain response of free standing monolayer
graphene has been found by Lee {\it et al}~\cite{Changgu} in pushing down a rigid tip of atomic force microscopy (AFM) which is placed over graphene. They determined that graphene is
a very strong material as much strong as diamond. Furthermore, based on experimental measurements, a linear slope of force-displacement curves (FDCs) has been measured.~\cite{frank,scott}

The physical scales available to computer simulations strongly depend on the model. In general, the level of detail that the model encompasses reduces both the number of atoms and length of times that can be simulated. A sophisticate approach is the density functional theory~\cite{dft} where the electronic structure of the atoms is calculated self-consistently and it can generally treat only on the system incorporate in order of several hundred atoms. Classical Molecular Dynamics simulation (MD), on the other hand is a common method to do several physical calculations for nano-indentation on various materials~\cite{thesis,diamondjournal} to analysis systems containing huge number of particles with time scales on the order of nanoseconds. Accordingly, we use MD to study nano-indentation of a circular graphene flake (CGF) and to the best of our
knowledge there is no any systematic study of simulated nano-indentation of
the CGF.

In experiment, since
the hardness of AFM's tip is required the diamond tip material is typically used. In general, geometrical properties of the three dimensional indenter on a bulk material are important either in experiments or simulations of nano-indentation.~\cite{thesis} In order to check whether or not the existence of a geometric dependency, we examined two different structures, the diamond structure with carbon atoms and a face-centered cubic  structure incorporating Platinum atoms, for an indentation simulation and found no indication of geometrical dependence in two dimensional graphene flake. In other words, our results indicate that the results are independent of the geometry of indenter at least in the two structures considered for the indentation problem in graphene.

The deformation of CGF under a central point load by carrying out molecular mechanics simulations has been investigated.~\cite{duan} In the recent work, authors used properly selected parameters and showed that the von K$\acute{a}$rm$\acute{a}$n plate theory can provide a remarkable accurate prediction of the graphene sheet behavior under linear and nonlinear bending and stretching. In the present letter, we have used the theory of elasticity based on
F\"{o}pple-Hencky equations for a CGF and shown that the theory is applicable to graphene sheets by comprising the results of theory with those results calculated by MD simulations.

\begin{figure}
\includegraphics[width=8.5cm]{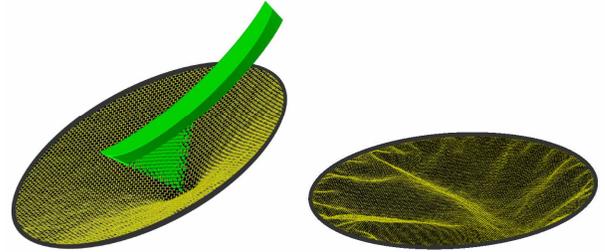}
\caption{(color online). Left panel: A snapshot of an indenter is placed over a roughed graphene
circular flake with cramped boundary conditions at room temperature. Right panel: A snapshot of circular graphene flake under influence of boundary tension before loaded by the indenter. }
\end{figure}

\noindent
{\em Theory and Model}---
We have used MD method to
simulate nano-indentation of a suspended CGF. Number of
carbon atoms in our simulation varies from $18740$ to $110096$
correspond to graphene surfaces with radius $R=12$ to
$R=30.1$ nm, respectively. A rigidly clamped boundary condition has been
employed. In our simulations, the indenter is composed of $371$ atoms. We
have simulated the system at room temperature, ($300 K$) by employing
Nos'e-Hoover thermostat.
 The Brenner's bond-order potential~\cite{brenner, neek-amal} has been
 used for carbon-carbon interactions and Van der-Waals potential
for the indenter-graphene interactions~\cite{erkoc2001}. For a two-component system, as
studied here, the parameters for the mixed interaction between
two type of atoms can be estimated by the simple average
suggested by Steel {\it et al}~\cite{steel}. In the beginning of
simulation, the lowest position of tip's atoms is considered a few angstroms, i.e. $\approx 2.7$
${\AA}$ above graphene sheet. In the load process, the indenter is pushed down slowly as $0.1 \AA$ over
$5000\Delta t$ which is equivalent to a velocity equal to $4$ m/s. To avoid unphysical effects due to step time, the length of moving indenter steps was chosen to be small with respect to the force cutoff length for the interatomic potential. The indenter
atoms interact only with graphene atoms via the Lennrad-Jones
interactions. The shape of the indenter was chosen as a square-based pyramid as shown in Fig.~1.

We have also applied the theory of elasticity~\cite{landau} for the CGF
under concentrated force at large
deflection, $z$. Note that a typical size of system
is much larger than the deflection values, and in addition the thickness
of flake is much smaller than deflection amounts.

The energetics considerations of a graphene in the limit of large
deflections include both bending and stretching energies.~\cite{landau} The condition of minimum energy for graphene flake yields the F\"{o}ppl equation. The solutions of equations can be given by
using the Hencky transformations. The governing equations in
planar-polar coordinate are written as follow
\begin{eqnarray}\label{Eq1}
r\frac{d}{dr}[\frac{d}{r dr}(r^2 \sigma_{r})]&=&-\frac{t E}{2}(\frac{dz}{dr})^2,
\nonumber\\\sigma_{r} \frac{dz}{dr}&=&-\frac{F}{2\pi r}
\end{eqnarray}
 where $r$ is the radial
position, $R$ is the radius of CGF, $z(r)$ is the deflection, $t$ is the
thickness of graphene approximated by 1 {\AA}, $E$ is the Young's modulus, $\sigma_{r}$
is the radial stress of flake and $F$ is the concentrated load on the
flake. We are interested on rigidly
clamped boundary condition or fixed boundary condition in the
absent of residual stress, i.e., $z=0$ at $r=R$ and it impose the following equation~\cite{landau}
\begin{equation}\label{Eq2}
R\frac{d\sigma_{r}}{dr}+(1-\nu)\sigma_{r}=0~,
\end{equation}
where $\nu$ is Poisson's ratio. Expressions given by Eq.~(\ref{Eq1}) are nonlinear equations. At large deflection value, the set of equations can be solved analytically~\cite{cong} with considering Eq.~(\ref{Eq2}) and the concentrated force leads $F=\frac{\pi E t}{4R^2}\frac{1}{G(\nu)}\zeta^3$
where $G(\nu)$ is a complicated function of the Poisson ratio. Here $\zeta=|\overrightarrow{\zeta}|$ where $\overrightarrow{\zeta}=-z(0)\hat{k}$ is graphene deflection at the center of CGF, $r=0$ in the $\hat{k}$ direction. $G(\nu)$ has almost a linear
behavior with respect to Poisson's ratio of desired systems and mostly
 varies from $0.9$ to $0.65$ for given $\nu\simeq0.0$ to $0.5$. Surprising enough, our simulation results confirm
the concentrated force behavior which we will discuss that in discussion Section.
\begin{figure}[t]
\includegraphics[width=6.6cm]{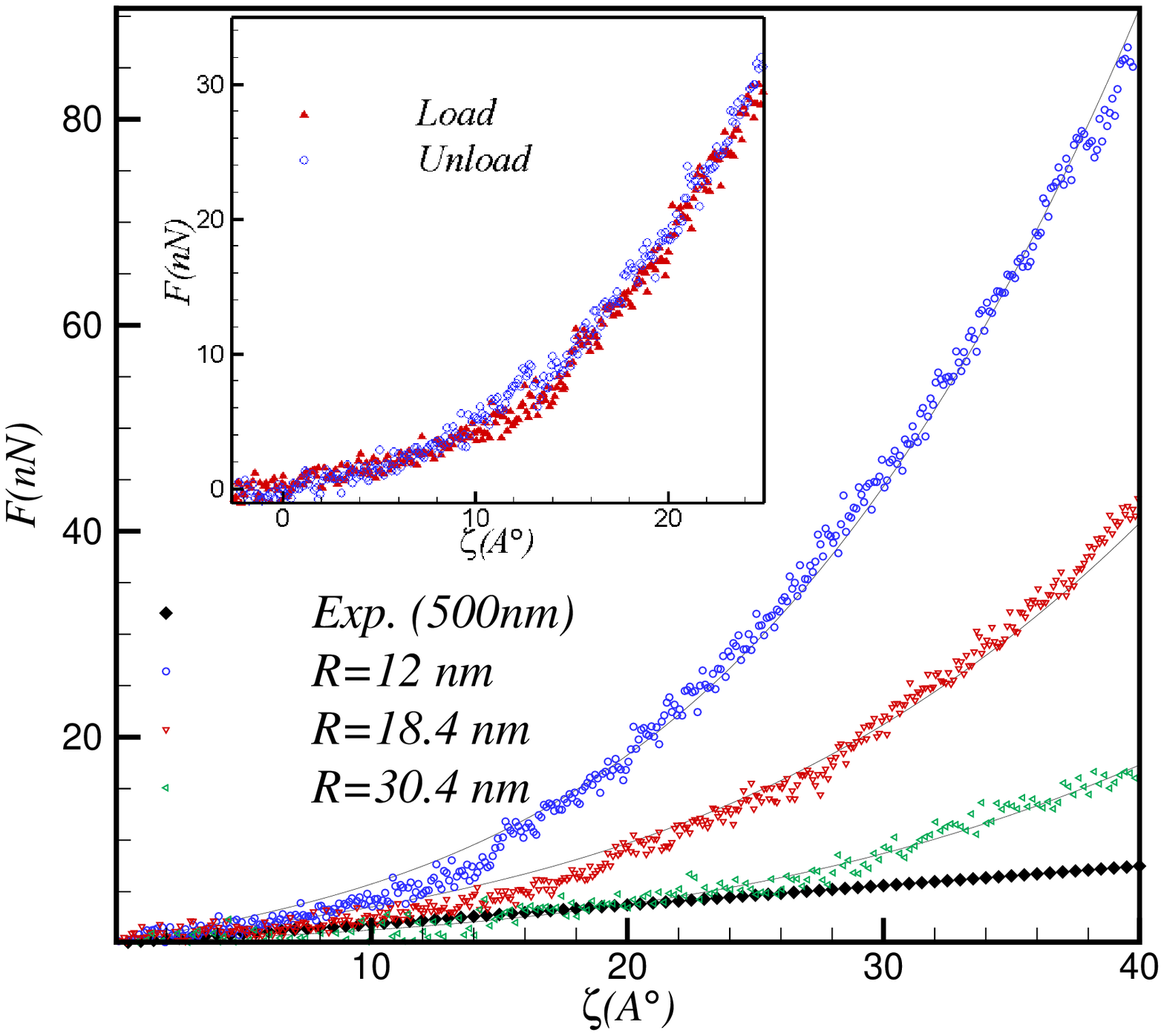}
\includegraphics[width=5cm]{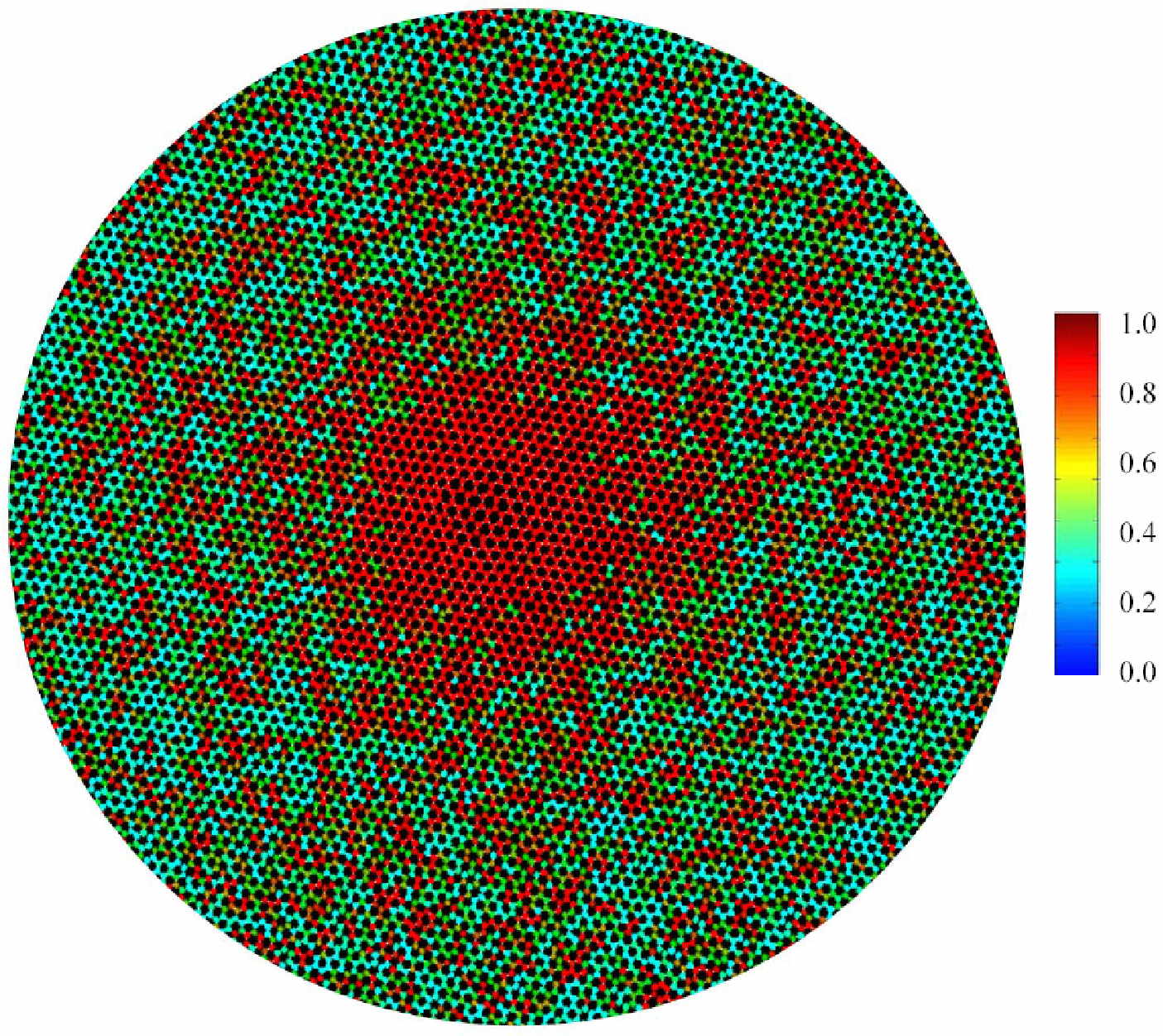}
\caption{(color online). Top: Load force as a function of displacement.
The fitting function given by Eq.~(\ref{Fexp}) for three different sizes are
shown by solid curves. Filled symbols are experimental data, Ref [\onlinecite{Changgu}]. In the Inset: Load and unload ( the indenter pulled up from maximum deflection position at $\zeta=27$ nm ) force as a function of displacement for the graphene flake with $R=12$ nm. Bottom: Distribution of stress on graphene flake in arbitrary units when the indentor was at $\zeta=27$ nm. }
\end{figure}

The mechanical properties of the free standing graphene
can be probed by the tip of AFM. The experimental results have
obtained for deflections greater than $10 nm$. Recently, the
FDCs have been measured by Lee {\it et al}~\cite{Changgu} and they have shown that the FDCs can be approximated by a simple
polynomial function having a linear term due to tension effect and a cubic term due to bending effect,
\begin{equation}\label{Fexp}
F=a\zeta+b\zeta^3,
\end{equation}
and thus the bending term will dominate at large deflection. The radius dependence of Hooke coefficient can be determined from the theory of elasticity~\cite{safran} in which the Euler-Lagrange (EL) equation of minimization of the free energy is $-\sigma \nabla^2 z(\overrightarrow{x})+\kappa \nabla^4 z(\overrightarrow{x})=F \delta(\overrightarrow{x}-\overrightarrow{x_0})$ where the $\sigma$ and $\kappa$ are respectively the tension and the bending rigidity and $\overrightarrow{x}=(x,y)$. Accordingly, the Hooke coefficient of Eq.~(\ref{Fexp}) has a complicated dependence on
$R$. We have $z(q)=F/(\sigma q^2+\kappa q^4)$ from the EL equation and it leads that $a\sim -4\pi \sigma/\ln(\sigma/(\sigma R^2+\kappa))$ . In this work, we therefore try to find a relationship between the parameter $b$ with the radius of CGF.

\noindent
{\em Discussions}---
Figure~1 (left panel) shows a snapshot of simulated CGF with
an indenter placed over graphene (the arm of tip is shown only for clarity). The $z$-component of forces from graphene atoms on indenter are calculated by summing over total
reaction forces. Note that the deflection or displacement $z$, of
CGF is equivalent to the indenter displacement in each
step. It is essential to compare the simulation results of FDCs with those measured in experiments for the cramped CGF.



\begin{figure}
\includegraphics[width=7cm]{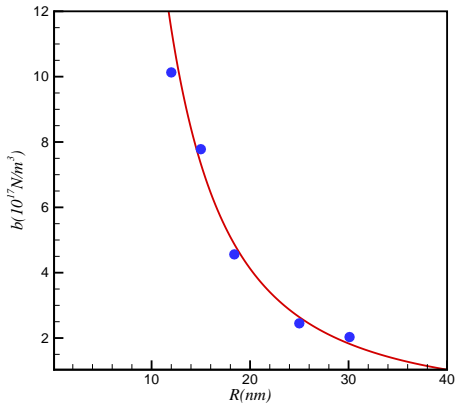}
\caption{(color online). Values of parameter $b$ as a function of graphene circular size denoted by symbols. The solid line shows a $1/R^2$ function.}
\end{figure}

Figure~2 (top) shows the variation of applied load at $r=R$ as a
function of graphene's deflection in $z$-direction. Corresponding
values for parameters $a$ and $b$ have been obtained by fitting the numerical results with the expression given by Eq.~(\ref{Fexp}). The size dependence of FDCs is determined in the figure. It is worthwhile to note that the sample size in the experiment possessing radius $500$ nm is much larger than the one in our simulation samples. The FDC results for $R=30.1$ nm where $a=0.108$ N/m and $b=2.03\times10^{17}$N/m$^3$, are very close to the data of experiments. Note that there are defects and impurities in a real sample and those would reduce the FDCs values.
Apparently, studying the indentation properties for either a tiny graphene
sheet or a tiny tip should be complicated in experiment. Obviously, the physics of indentation for tiny graphene sheet can be understood based on our MD simulations.

One important physics which can be extracted from our simulation is unload FDCs.
In the unload case where graphene was
already loaded and reached to a maximum deflection value where $\zeta=27$ nm and then the indenter is pulled up, we have found that the force result in this case is the same as it in loaded case. In the inset of Fig.~2, the force results as a function of displacement for both load and unload cases are shown. It is physically known that the load and unload FDCs would be different when a bulk material is used. The physical reason is the deformation of structure make in the unload process. The distribution of stress~\cite{landman} on the CGF which depends on a mass density, Young's modulus and Poisson's ratio is shown in Fig.~2 (bottom) as well. The large stresses generated near the indenter at the center and decays fast as a function of radial value since the interaction of carbon-carbon is nonharmonic in the model interaction. In addition, the distribution of stress is no longer uniform on the surface due to the thermal fluctuations.

In order to get quantitative results, we compare the concentrated force calculated from the theory and simulation results fitted by Eq.~(\ref{Fexp}) to describe the radial dependence of parameter $b$. The parameter $b$ is given by $b=\frac{\pi E t}{4G(\nu)}\frac{1}{R^2}$.
We have obtained the value of $b$ by simulating many CGF sizes and results are shown in Fig~3 as a function of $R$. The value of $b$ decreases by increasing $R$ and it is fitted very well to $1/R^2$ function ( the slop of line in log-log plot is quite close to -2). It obviously means that the theory of elasticity is applicable for graphene flakes. Furthermore, we obtained that $\pi E t/ G(\nu)\approx 4\times165$ (N/m) using the fitting coefficient. Using approximated values
for $t\approx 1.0 \AA$ and $G\approx0.8$, the Young modules can be obtained and it is about $1.65$ Terra Pascal which is very close to the value predicted by recent experiments and theoretical calculations~\cite{Changgu,scott, fasolino}. Moreover, the $\sigma$ and $\kappa$ are respectively obtained about $0.16$ N/m and $0.8$ eV from the fitting values of Hooke coefficient which are very close to values have been predicted for graphene.~\cite{Giem2008, tu}


Interestingly, the mechanical vibration in suspended few-layer graphene has been illustrated in experiment by using a scanning prob microscopy and found a resonance frequency about $31$ MHz.~\cite{garcia} From the simulation point of view, when the indenter is pulled upward from its initial position there will be a position, ($z(R)\hat{k}$) where graphene sheet separates from the indenter (after the range of interaction between the carbon atoms and the indenter atoms) and the surface vibrates with high frequency. We have obtained the size dependence of frequency for one layer graphene and $\nu=32.0\pm0.1$, $19.0\pm0.3$ and $15.5\pm0.45$ in units of GHz for $R=12$, $15$ and $18.4$ nm, respectively shown that $\nu$ decreases by increasing $R$. We monitor that the vibration frequency behaves as a sinusoidal manner at small CGF size, ($12$ nm). Our finding regard to the frequency at small graphene size would be verified by experiments.

In the other hand, graphene under influence of boundary tension (GIBT) has been studied in several experiments.~\cite{meyer, Changgu} It would be worth to explore the FDCs behavior of GIBT. To this purpose, we generated a GIBT by
compressing the boundary of CGF about $1.2\%$ of its initial diameter and then the indenter over the GIBT is loaded with cramping boundaries condition.

We have shown a snapshot of a GIBT in Fig.~1 (right panel). The out-of plane
height on the surface is around $18 {\AA}$ which is much bigger than the CGF where extra tensions are zero. The results of FDCs
for the both cases have been shown in
Fig.~4 for $R=18.4$ nm. Apparently, the force value of GIBT is smaller than one calculated for the CGF. The reason is that the number of atoms in the board range of interaction potential in the GIBT is larger than the one in the CGF and hence the attractive part of total force increases. Accordingly, the total force in this case might be smaller than the one calculated in the CGF. Consequently, the effect of boundary tension is essential for studying FDCs in comparison with the experiment measurements.
\begin{figure}
\includegraphics[width=7cm]{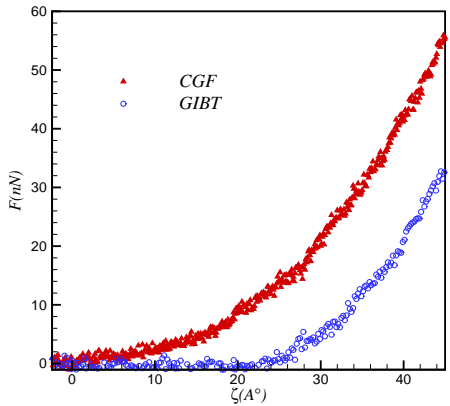}
\caption{(color online). Load force as a function of displacement for both CGF (red triangles) and GIBT (blue circles). }
\end{figure}

In summary, our simulation results show that the FDCs is almost
independent of ordinary temperatures( up to 400 K) and they
are close to the experimental data. Moreover, we found that the force on indenter in the CGF is larger than the GIBT. Importantly, the vibration frequency of the CGF in order of GHz is calculated and shown that it decreases by increasing the CGF sizes and essentially a particular vibration having a resemblance to a sinusoidal behavior occurs at small $R$. Eventually we would like to emphasize that our numerical results show that the continuum elasticity theory can be used for the indentation problem in graphene.

We gratefully acknowledge valuable comments and discussions from F. Guinea, F. Peeters, A. Naji. M. N was supported by Sh. R. Univ. under grant program.

\pagebreak

\begin{thebibliography}{99}
\bibitem{novoselov}
    K. S. Novoselov, A. K. Geim, S. V. Morozov, D. Jiang, Y. Zhang, S.
    V. Dubonos, I. V. Grigorieva, and A. A. Firsov, Science {\bf 306}, 666 (2004)~.
\bibitem{geim}
    A. K. Geim and K. S. Novoselov, Nature Mater. {\bf 6}, 183 (2007)
    ; A. H. Castro Neto, F. Guinea, N. M. Peres,
     K. S. Novoselov and A. K. Geim, Rev. Mod. Phys. {\bf 81}, 109 (2009)~.
\bibitem{Giem2008}
    Tim J. Booth, Peter Blake, Rahul R. Nair, Da Jiang, Ernie W. Hill, Ursel Bangert, Andrew Bleloch, Mhairi Gass, Kostya S. Novoselov, M. I. Katsnelson and A. K. Geim , Nano Lett. {\bf 8}, 2442 (2008)~.
\bibitem{Changgu}
    Changgu Lee, Xiaoding Wei, Jeffrey W Kysar and James Hone, Science {\bf 321} ,385 (2007)~.
\bibitem{frank}
    I. W. Frank, D. M. Tanenbaum, A. M. van der Zander and P. L. McEuen, J. Vac. Sci. Technol. B {\bf 25} ,2558 (2007)~.
\bibitem{scott}
    J. Bunch {\it et al.}, Science {\bf 315}, 490 (2007); J. Scott Bunch {\it et al.},  Nano. Lett. {\bf 8} ,2458 (2008)~.
\bibitem{dft}
    W. Kohn and L. J. Sham, Phys. Rev. A {\bf 140}, 1133 (1964)~.
\bibitem{thesis}
Jin-Yuan Hsieh, {\it et al.} Nanotechnology, {\bf 18}, 415701 (2007);
    Francis Brent Neal, Ph.D Thesis, Dept. Phys and Astr. Phys. North California University (2008).
\bibitem{diamondjournal}
    A. Richter, R. Ries, R. Smith, M. Henkel and B. Wolf, Diamond and related materials  {\bf 9}, 170 (2000)~.
\bibitem{duan}
    W. H. Duan and C. M. Wang, Nanothechnology {\bf 20}, 075702 (2009)~.
\bibitem{brenner}
    D. W. Brenner, Phys. Rev. B {\bf 42}, 9458 (1990)~.
\bibitem{neek-amal}
    N. Abedpour, M. Neek-Amal, R. Asgari, F. Shahbazi, N. Nafari, and
    M.R. Tabar, Phys. Rev. B {\bf 76}, 195407 (2007); M. Neek-Amal, R. Asgari and M. R. Rahimitabar, Nanotechnology {\bf 20}, 135602 (2009)~.
\bibitem{erkoc2001}
    S. Ercok, Ann. Rev. Comp. Phys. IX {\bf1}, 103(2001)~.
\bibitem{steel}
    H. A. Steel {\it The Introduction of Gases with Solid Surfaces},
    Oxford, Pergamon (1974)~.
\bibitem{landau}   L. D. Landau , L. P. Pitaevskii , A. M. Kosevich
    , E.M. Lifshitz {\it Theory of Elasticity, Third Edition}.
\bibitem{cong} JIN Cong-rui, Appl. Math. Mech. -Engl.
    Ed. {\bf 29}, 889 (2008)~.
\bibitem{safran}
    Samuel Safran {\it Statistical Thermodynamics of Surfaces, Interfaces, and membranes} (Addison-Wesley Publishing Company) (1994);  Eun-Ah Kim, A. H. Castro Neto, Europhysics Letters {\bf 84}, 57007 (2008)~.
\bibitem{landman}
U. Landman, W. D. Luedtke, J. Ouyang, T. K. Xia, Jpn. J. Appl. Phys. {32}, 1444 (1993)~.
\bibitem{fasolino}
    K. V. Zakharchenko, M. I. Katsnelson and A. Fasolino, Phys. Rev. Lett. {\bf 102}, 046808 (2009)~.
\bibitem{garcia}
    D. Garcia-Sanchez, et al, Nano Letter {\bf 8}, 1399 (2008)~.
\bibitem{tu}
    C. Z. Tu and Z. C. Ou-Yang, Phys. Rev. B{65}, 233407 (2002)~.
\bibitem{meyer}
    J. C. Meyer, A. K. Geim, M. I. Katsnelson, K. S. Novoselov, T.
    J.Booth and S. Roth, Nature {\bf 446}, 60 (2007);
    M. Ishigami, J.H. Chen, W.G. Cullen, M.S. Fuhrer, and E.D.
    Williams, Nano Lett. {\bf 7}, 1643 (2007)~.


\end{thebibliography}
\end{document}